\documentclass[aps,prl,twocolumn,superscriptaddress,longbibliography]{revtex4-2}

\usepackage{mathrsfs,amssymb,graphics,subfigure,threeparttable}
\usepackage{graphicx}
\usepackage{amssymb}
\usepackage{enumerate}
\usepackage{amsmath}
\usepackage{fancyhdr}
\usepackage{bm}
\usepackage[squaren]{SIunits}
\usepackage{xcolor}
\usepackage{changes}
\usepackage{soul}
\usepackage{multirow}
\usepackage{hyperref}
\hypersetup{
    colorlinks=true,
    linkcolor=blue,
    filecolor=black,      
    citecolor=blue,
    urlcolor=blue,
    }

\begin{document}

%\preprint{}

%\title{High brightness, low energy spread relativistic beam generation in a self-injected, self-loaded, high-efficiency plasma wakefield accelerator stage}

\title{Damping dynamics of the centroid oscillation of a relativistic laser pulse in a plasma channel}

\author{Yuhui Xia}
\author{Zhenan Wang}
\author{Ziyao Tang}
\author{Jianghao Hu}
\author{Xinyang Liu}
\author{Letian Liu}
\author{Laifu Man}
\author{Zhuo Pan}
\author{Di Wu}
\affiliation{State Key Laboratory of Nuclear Physics and Technology, and Key Laboratory of HEDP of the Ministry of Education, CAPT, Peking University, Beijing 100871, China}
\author{Jacob R. Pierce}
\affiliation{Department of Physics and Astronomy, University of California Los Angeles, Los Angeles, California 90095, USA}
\author{Xueqing Yan}
\affiliation{State Key Laboratory of Nuclear Physics and Technology, and Key Laboratory of HEDP of the Ministry of Education, CAPT, Peking University, Beijing 100871, China}
\affiliation{Beijing Laser Acceleration Innovation Center, Beijing 100871, China}
\affiliation{Guangdong Institute of Laser Plasma Accelerator Technology, Baiyun, Guangzhou, 510540, China}
\author{Chen Lin}
\author{Xinlu Xu}
\email{xuxinlu@pku.edu.cn}
\affiliation{State Key Laboratory of Nuclear Physics and Technology, and Key Laboratory of HEDP of the Ministry of Education, CAPT, Peking University, Beijing 100871, China}
\affiliation{Beijing Laser Acceleration Innovation Center, Beijing 100871, China}

\date{\today}

\begin{abstract}
The centroid oscillation of an offset laser pulse propagating in a preformed plasma channel is investigated through theoretical analysis and three-dimensional particle-in-cell simulations. For non-relativistic laser pulses, the mode leakage of a finite channel and the temporal walk-off between the fundamental and high order modes of a finite-duration laser induce a decay in the laser centroid oscillation. An analytical model characterizing these decay mechanisms is derived and validated by simulations. For relativistic laser pulses, the slice-based centroid oscillation frequency develops an axial chirp due to relativistic channel modification and photon deceleration. This chirp leads to phase mixing across different axial slices of the pulse, resulting in a rapid damping of the overall centroid oscillation. Understanding this oscillation damping is crucial for mitigating electron beam pointing jitter and maintaining beam quality in high-energy, channel-guided laser wakefield accelerators.
\end{abstract}

%\pacs{}
%\keywords{}

\maketitle
%%%%%%%%

%\section{}
%\subsection{}
%\subsubsection{}

\section{\label{sec:level1}I. Introduction}
An ultraintense and ultrashort laser pulse can drive a nonlinear plasma wave wake with $\sim\giga\volt\per\centi\meter$ acceleration gradients in an underdense plasma \cite{PhysRevLett.43.267} and thus holds the promise to shrink the size and reduce the cost of conventional particle accelerators \cite{faure2004laser, geddes2004high, mangles2004monoenergetic}. Over the past decade, laser wakefield acceleration (LWFA) has achieved major milestones, including generation of $\sim 10$ GeV electron beams in plasma only 10s of centimeters long \cite{aniculaesei2024acceleration, picksley2024matched}, production of GeV-class high-quality electron beams \cite{wang2016high, ke2021near}, and using such bright beams to drive free-electron lasers (FELs) \cite{wang2021free, labat2023seeded,vh62-gz1p, Jin2026PRR, Jiang2026PRR}. These advances have prompted serious consideration of LWFA as potential building blocks for next-generation TeV-class linear colliders \cite{Schroeder_2023}.

In LWFA, the laser pulse is usually tightly focused to spot sizes of a few 10s of microns to be commensurate with the plasma wavelength. External guiding \cite{Durfee1993PhysRevLett.71.2409, leemans2006gev, leemans2014multi, PhysRevLett.122.084801, Miao2022PhysRevX.12.031038, Picksley2023PhysRevLett.131.245001, Shrock2024PhysRevLett.133.045002, picksley2024matched, Lahaye2025PRAB} and/or self-guiding \cite{hafz2008stable, PhysRevLett.102.175003, Ralph2010PoP, PhysRevLett.103.035002, PhysRevLett.105.105003, PhysRevLett.107.035001, PhysRevLett.107.045001, wang2013quasi, PhysRevLett.111.165002, ke2021near, aniculaesei2024acceleration} is therefore necessary to maintain a stable laser spot size and sustain the plasma wake over many Rayleigh lengths needed for GeV-class acceleration. A preformed plasma channel with a transverse density profile of $n_\mathrm{p}(r)=n_\mathrm{p0} + \Delta n\frac{r^2}{w_\mathrm{ch}^2}$ has been widely adopted in high-energy LWFA to guide the laser pulse, where $n_\mathrm{p0}$ is the on-axis density, $\Delta n=\frac{1}{\pi r_e w_\mathrm{ch}^2}$ is the channel depth and $w_\mathrm{ch}$ is fundamental channel mode radius, $r_e\approx 2.82\times 10^{-15}~\meter$ is the classical electron radius. This parabolic profile can provide ideal guiding for a nonrelativistic laser pulse with a transversely Gaussian intensity profile when the spot size $w_0$ is equal to $w_\mathrm{ch}$. However, in reality, the laser may enter the plasma channel with a mismatched spot size, or with a transverse position and/or angle offsets relative to the channel axis. These practical imperfections, combined with the relativistic intensity of the laser pulse, can complicate the subsequent laser propagation, distort the plasma wake, and degrade the quality of the accelerated beam.

If the laser spot size is mismatched to the channel ($w_0\neq w_\mathrm{ch}$), it can couple into multiple high order guided modes. The phase velocity difference between these modes and the fundamental mode leads to mode beating, which manifests as oscillations of the peak laser intensity \cite{FederMiao2020PhysRevResearch.2.043173}. The oscillation  vanishes for ultrashort pulses as the beating modes walk off temporally due to the group velocity dispersion \cite{EsareyPhysRevE.59.1082, FederMiao2020PhysRevResearch.2.043173}. Lasers with relativistic intensity can modify the channel profile and make the mode beating more complex \cite{Miao2022PhysRevX.12.031038, Picksley2023PhysRevLett.131.245001, Shrock2024PhysRevLett.133.045002}. The concept of a supermatched laser pulse with a slice-dependent intensity distribution is proposed to account for the laser-induced channel distortion and enable stable propagation \cite{Benedetti2012PoP, Benedetti2015PhysRevE.92.023109}. Plasma channels with finite thickness, such as these formed via hydrodynamic optical field ionization (HOFI) \cite{Durfee1993PhysRevLett.71.2409}, support quasi-bound or leaky modes, i.e., the intensity of the mode decreases exponentially along the channel \cite{Clark1998PhysRevLett.81.357, ClarkPhysRevE.61.1954}. The fundamental mode typically has a longer attenuation length than high order modes. This effect can be leveraged as a mode filter, radiating away the deleterious high order modes and guiding the fundamental mode for LWFA \cite{Antonsen1995PhysRevLett.74.4440, ClarkPhysRevE.61.1954, picksley2024matched}.

If a non-relativistic laser pulse enters a parabolic plasma channel with a transverse offset and/or an angular misalignment, its centroid oscillates with a constant amplitude and a period of $2\pi z_\mathrm{R}$ \cite{Gonsalves2010PoP}, a phenomenon that has been exploited to realize a plasma undulator \cite{Rykovanov2015PhysRevLett.114.145003, rykovanov2016tunable}, where $z_\mathrm{R}=\frac{\pi w_0^2}{\lambda_\mathrm{L}}$ is the Rayleigh length and $\lambda_\mathrm{L}$ is the laser wavelength. In LWFA-relevant scenarios, the centroid motion is further influenced by leakage from channels of finite width \cite{Miao2022PhysRevX.12.031038}, walk-off between different modes and relativistic laser–plasma interactions. Such laser centroid oscillations can wiggle the accelerated electron beam, introduce beam-pointing jitter, and may even seed instabilities that degrade beam quality \cite{Picksley2023PhysRevLett.131.245001}, making their understanding critical for future applications, e.g., x-ray FELs and colliders. 

In this work, we use theoretical analysis and particle-in-cell (PIC) simulations to study the motion of the laser centroid in LWFA and find the centroid oscillation amplitude decays for both non-relativistic and relativistic laser pulses. The amplitude decay in the nonrelativistic case is caused by the mode leakage of a finite channel and the temporal walk-off between the fundamental and high order modes of a finite-duration pulse. An analytical model incorporating both effects is given and validated via simulations. The relativistic laser pulse deepens the plasma channel, resulting in an axially varying matched spot size. Meanwhile, photon deceleration induces a frequency chirp along the laser pulse. These two effects cause a decreasing oscillation period from the laser head to the tail, leading to phase mixing of the centroid oscillation at different axial slices. A rapid decay of the averaged laser centroid oscillation is thus present. A series of simulations is performed for different channel widths and $a_0$ values to validate the analysis. Understanding the centroid dynamics of an offset laser in a plasma channel is essential for high-quality and high-energy LWFAs and the damping of the oscillation may be beneficial for reducing the pointing jitter of the accelerated electrons.

\section{\label{sec:level1}II. Results from particle-in-cell Simulations}
\begin{figure}
\centering
\includegraphics[width=1.0\linewidth]{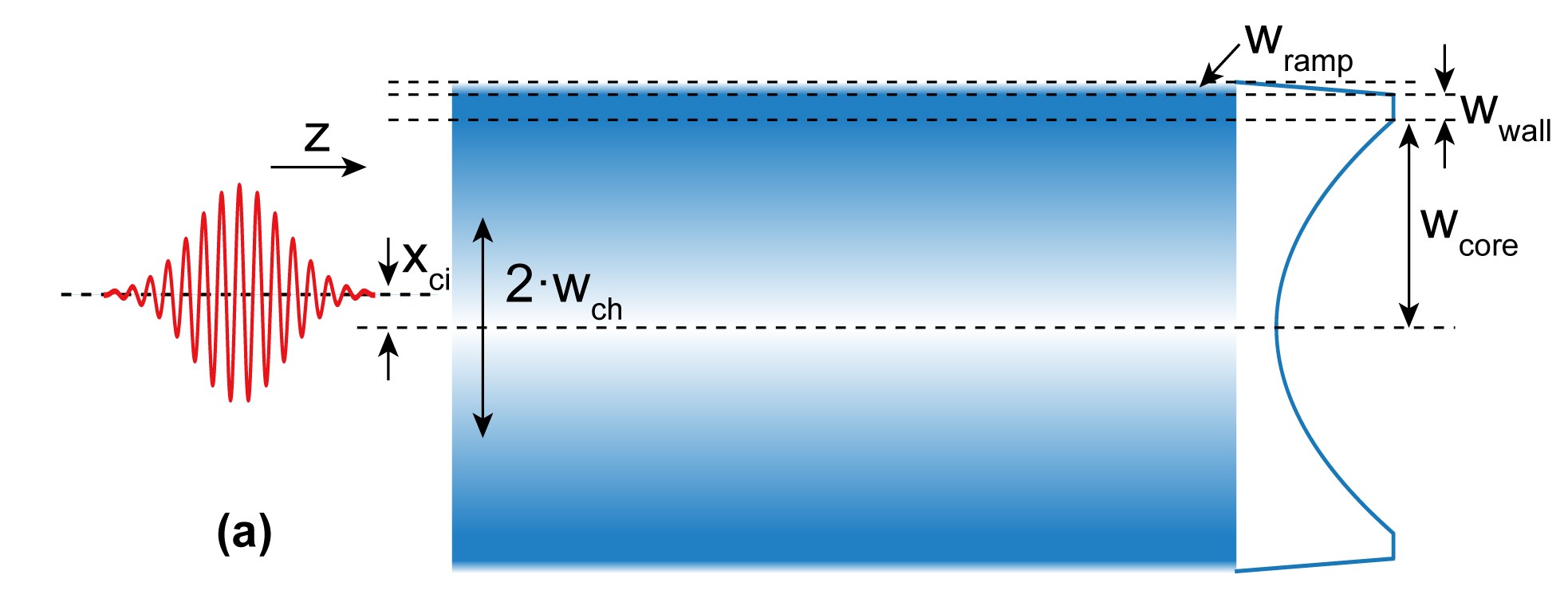}
\includegraphics[width=1.0\linewidth]{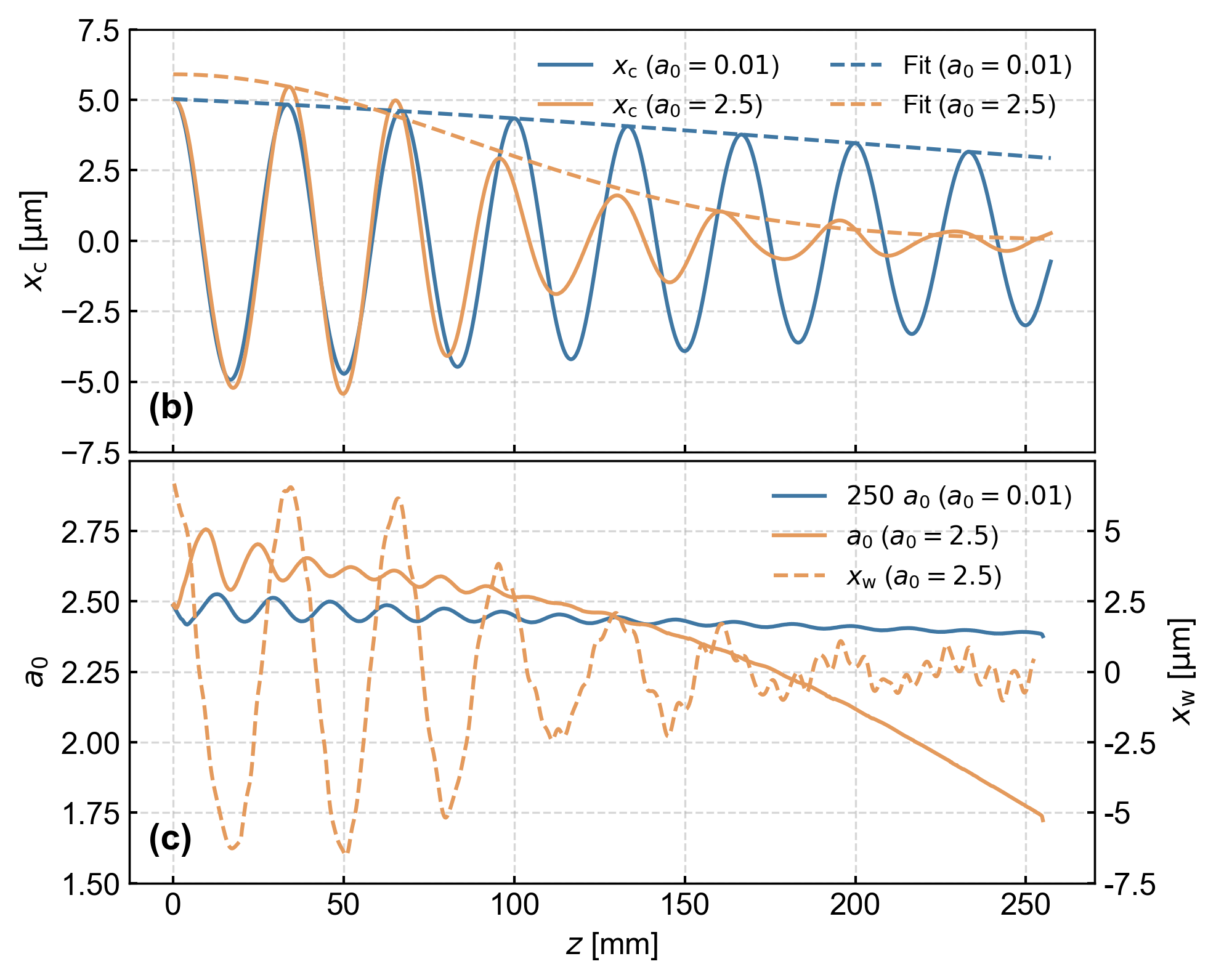}
\caption{\label{fig:mainresults} (a) Schematic of a laser pulse propagating in a plasma channel with an initial misalignment. The oscillation of the laser centroid (b) and the evolution of the peak $a_0$ (c) in simulations. The blue lines in panels (b) and (c) correspond to $a_0 = 0.01$, whereas the orange lines correspond to $a_0 = 2.5$. The blue dashed line in (b) represents an envelope fit based on Eq. \eqref{eq:D} and the orange dashed line shows a Gaussian fit. The orange dashed line in (c) indicates the oscillation of the center of the first wake when $a_0=2.5$.}
\end{figure}

As shown in Fig. \ref{fig:mainresults}(a), a relativistic linearly polarized (LP) laser pulse propagates in a preformed plasma channel with an initial transverse position or angle offset. We consider collider-relevant parameters \cite{Schroeder_2023}. Current design of LWFA-based colliders consists of $\sim$100 stages, with each stage boosting the electron beam by $\sim 10$ GeV over plasmas that are tens of centimeters long \cite{Schroeder_2023}. A 1-$\micro\meter$ wavelength laser pulse with a spot size of $w_0=41\micro\meter$ and a FWHM duration of $\tau_\mathrm{L}=42$ fs propagates in a plasma channel with $n_\mathrm{p0}=9\times 10^{16}~\centi\meter^{-3}$ and $w_\mathrm{ch}=w_0$, and the plasma length is $L_\mathrm{p}=25.5~\centi\meter$ \cite{Schroeder2025FLC}. The peak normalized vector potential of the laser is $a_0=\frac{eE_\mathrm{L}}{mc\omega_\mathrm{L}}=2.5$ and the peak power is 0.22 PW, where $E_\mathrm{L}$ and $\omega_\mathrm{L}$ are the peak electric field and the angular frequency of the laser, respectively. The laser is focused at the entrance of the plasma channel. We perform three-dimensional (3D) PIC simulations using fully relativistic code OSIRIS \cite{fonseca2002high} to investigate the motion of the laser centroid in the presence of a misalignment. The simulation setup can be found in the Appendix A. The simulations are performed in a boosted frame with $\gamma_\mathrm{b}=10$ to save the computational cost \cite{yu2014modeling, yu2016enabling}. An advanced Maxwell solver \cite{li2017controlling, xu2020numerical} is used to eliminate the numerical noise related with relativistically drifting plasma in the boosted frame \cite{xu2013numerical, xu2020numerical}.

Complex amplification and compression procedures of the ultrashort and high-power laser systems, with long-beam paths and numerous optical components, induce many fluctuations to the laser properties at focus \cite{Danson2015HPLSE, Danson2019HPLSE}, such as position and pointing jitters \cite{isono2021high}. For example, the BELLA 100 TW laser with an $f=3~\meter$ off-axis parabolic mirror achieves a position jitter of $\sim5~\micro\meter$ and a pointing jitter of $\sim 200~\micro\radian$ at the focus \cite{isono2021high}. An active stabilization system based on a nonperturbative diagnostic for the high-power laser delivery at focus has been developed to reduce the jitters \cite{Berger2023PhysRevAccelBeams.26.032801, Amodio_2025}. We thus chose the initial offset of the laser centroid as $x_\mathrm{ci}=5~\micro\meter$, where $x_\mathrm{c}=\frac{\int \mathrm{d}x\mathrm{d}y\mathrm{d}t E_\mathrm{L}^2x}{\int\mathrm{d}x\mathrm{d}y\mathrm{d}t E_\mathrm{L}^2}$ is the position of the overall centroid and the subscript `i' indicates the initial quantities. An angular offset of $\theta_{x\mathrm{i}}$ is equivalent to a position offset of $z_\mathrm{R}\theta_{x\mathrm{i}}$ and we choose $\theta_{x\mathrm{i}}=0$ in the simulations without loss of generality.

%A representative example is the BELLA PW laser ($13.5\,\text{m}$ OAP), where machine-learning-based active stabilization significantly suppresses focal spot jitter from $\sigma_x = 13.1~\micro\meter$ and $\sigma_y = 14.8~\micro\meter$ (system off) to $\sigma_x = 4.6~\micro\meter$ and $\sigma_y = 7.9~\micro\meter$ (system on). We define an effective amplitude of laser centroid oscillation as $\sigma_\mathrm{eff} = \sqrt{\sigma^2 + (z_\mathrm{R}\theta)^2}$ to qualify the simulation parameters. In our case, the effective amplitude cased by the initial position offset of $5~\micro\meter$ is much larger than $Z_\mathrm{R}\theta=1.056~\micro\meter$ cased by the angular offset of $200~\micro\radian$.

A HOFI channel is capable of generating a low on-axis plasma density of $n_\mathrm{p0} \approx 10^{17}~\mathrm{cm}^{-3}$, satisfying the requirements for 10 GeV-class energy gain in collider applications \cite{picksley2024matched}. An alternative approach employs discharged capillaries with auxiliary laser heating \cite{PhysRevLett.122.084801}. However, thermal load and material erosion at high repetition rates pose significant challenges for collider applications. The HOFI channel is characterized by a finite parabolic core bounded by a high-density shock-wall based cladding \cite{Clark1998PhysRevLett.81.357, ClarkPhysRevE.61.1954}. We model the channel consisting of a parabolic plasma core, a flat wall, and a density downramp, with respective widths $w_\mathrm{core}$, $ w_\mathrm{wall}$ and $w_\mathrm{ramp}$ as shown in Fig. \ref{fig:mainresults}(a). The channel parameters are set to $w_\mathrm{ch}=41~\micro\meter$ and $w_\mathrm{core}=82~\micro\meter,~w_\mathrm{wall}=10~\micro\meter,~w_\mathrm{ramp}=5~\micro\meter$ in the simulations. 

The motion of $x_\mathrm{c}$ in the plasma channel is shown by the solid lines in Fig. \ref{fig:mainresults}(b) for two cases: $a_0=0.01$  (non-relativistic laser pulse) and $a_0=2.5$ (relativistic laser pulse). The centroid of the non-relativistic laser pulse (blue line) oscillates with a period of 33.2 mm. A slow decay of the oscillation amplitude ($\hat{x}_\mathrm{c}$) is present. As shown by the orange line, $\hat{x}_\mathrm{c}$ of the relativistic laser pulse decays much faster after the first two oscillation periods ($z\approx 70$ mm). Its oscillation period also gradually decreases relative to that of the non-relativistic pulse. 

The evolution of $a_0$ for the two cases is shown in Fig. \ref{fig:mainresults}(c) by the blue and orange solid lines, respectively. For the non-relativistic laser case, $a_0$ remains nearly constant during propagation, except for a small amplitude oscillation (the relative oscillation amplitude is $\sim 1\%$) with a period equal to half that of $x_\mathrm{c}$ (blue solid line). This oscillation is caused by the slight mismatch of the laser spot size and the transverse profile with the channel with a finite width. $a_0$ of the relativistic laser pulse decreases due to strong pump depletion (orange solid line) and it is 1.75 at the end of the plasma. The orange dashed line represents the motion of the first wake center (defined as the location where the wakefields satisfy $E_z=E_r=0$) in the relativistic case. It approximately follows the laser centroid oscillation.

\iffalse
\textbf{Case with a non-relativistic laser pulse} For an ideal and infinite plasma channel, the transverse eigenmodes can be equivalently formulated as Laguerre-Gaussian (LG) modes in cylindrical coordinates or Hermite-Gaussian (HG) modes in Cartesian coordinates. The complete orthonormal set of the HG modes is expressed as $u_n(x) = \left( \frac{2}{\pi w_\mathrm{ch}^2} \right)^{\frac{1}{4}} \frac{1}{\sqrt{2^n n!}} H_n\left(\frac{\sqrt{2}x}{w_\mathrm{ch}}\right) \exp\left(-\frac{x^2}{w_\mathrm{ch}^2}\right)$, where $H_n$ represents the Hermite polynomials of order $n$. For a Gaussian laser pulse with a matched spot size and an offset $x_\mathrm{c}$, its electric field can be decomposed as $E_\mathrm{L}(x,y)=  \hat{E}_\mathrm{L} \exp\left[-\frac{(x - x_\mathrm{c})^2 +y^2}{w_0^2}\right] = \sum_{n=0}^{+\infty}C_n u_n(x) u_0(y)$, where $C_n= \hat{E}_\mathrm{L} \left( \frac{\pi w_0^2}{2}\right)^{\frac{1}{4}}\frac{1}{\sqrt{n!}}  \exp\left(-\frac{x_\mathrm{c}^2}{2 w_0^2}\right) \left(\frac{x_\mathrm{c}}{w_0}\right)^n$. When $x_\mathrm{c}\ll w_0$, it suffices to retain only the fundamental Gaussian mode and the HG$_{10}$ mode \cite{Wang2017SR, LeiPhysRevAccelBeams.22.071302}, i.e., 
\begin{align}
    E_\mathrm{L}(x,y) \approx \hat{E}_\mathrm{L} \left( \frac{\pi w_0^2}{2}\right)^{\frac{1}{2}} \mathrm{e}^{-\frac{x_\mathrm{c}^2}{2 w_0^2}} \left[ u_0(x) + \frac{x_\mathrm{c}}{w_0} u_1(x) \right] u_0(y)
\end{align}
The centroid offset is directly proportional to the peak field ratio between the HG$_{10}$ and Gaussian modes. 
\fi

\section{III. Laser centroid oscillation in the non-relativistic case} 
For an ideal and infinite plasma channel, the transverse eigenmodes can be equivalently formulated as Laguerre-Gaussian (LG) modes in cylindrical coordinates $(r, \phi)$ or Hermite-Gaussian (HG) modes in Cartesian coordinates. The complete orthonormal set of the LG modes is expressed as $u_{p,m}(r,\phi) = \sqrt{\frac{2 p!}{\pi w_\mathrm{ch}^2 (p+|m|)!}} \left(\frac{\sqrt{2}r}{w_\mathrm{ch}}\right)^{|m|} L_p^{|m|}\left(\frac{2r^2}{w_\mathrm{ch}^2}\right) \exp\left(-\frac{r^2}{w_\mathrm{ch}^2}\right) e^{i m \phi}$, where $L_p^{|m|}$ represents the associated Laguerre polynomials with radial index $p$ and azimuthal index $m$. For a Gaussian laser pulse with a matched spot size ($w_0 = w_\mathrm{ch}$) and a transverse offset $x_\mathrm{c}$ along the $x$-axis, its electric field can be decomposed exclusively into the $p=0$ radial modes: $E_\mathrm{L}(r,\phi) = \hat{E}_\mathrm{L} \exp\left[-\frac{(x - x_\mathrm{c})^2 +y^2}{w_0^2}\right] = \sum_{m=-\infty}^{+\infty} C_{0,m} u_{0,m}(r,\phi)$, where the exact expansion coefficients are $C_{0,m} = \hat{E}_\mathrm{L} \sqrt{\frac{\pi w_0^2}{2}} \frac{1}{\sqrt{|m|!}} \exp\left(-\frac{x_\mathrm{c}^2}{2 w_0^2}\right) \left(\frac{x_\mathrm{c}}{\sqrt{2} w_0}\right)^{|m|}$. When $x_\mathrm{c} \ll w_0$, $C_{0,m}$ decreases rapidly with increasing $|m|$, and it suffices to retain only the fundamental Gaussian mode ($m=0$) and the first-order modes ($m=\pm 1$) \cite{Wang2017SR, LeiPhysRevAccelBeams.22.071302}, i.e.,
\begin{align}
    E_\mathrm{L} &\approx \hat{E}_\mathrm{L} \sqrt{\frac{\pi w_0^2}{2}} \mathrm{e}^{-\frac{x_\mathrm{c}^2}{2 w_0^2}} \left[ u_{0,0}+ \frac{x_\mathrm{c}}{\sqrt{2} w_0} \big( u_{0,1}+ u_{0,-1} \big) \right].
\end{align}
The centroid offset is directly proportional to the peak field ratio between the LG$_{0,\pm 1}$ modes and the fundamental mode. Note that we can also approximate the offset laser field as a superposition of the fundamental mode and the HG$_{1,0}$ mode. 

The wavenumbers of the eigenmodes are $\beta_{p,m}\approx k_\mathrm{L}\left[1- \frac{\omega_\mathrm{p0}^2}{2\omega_\mathrm{L}^2} - \frac{2(2p+|m|+1)}{k_\mathrm{L}^2w_0^2} \right]$. The difference in $v_{\mathrm{ph},p,m}=\frac{\omega_\mathrm{L}}{\beta_{p,m}}$ between the fundamental and the LG$_{0,\pm1}$ mode results in mode beating with a period of $\Lambda = \frac{2\pi}{\beta_{0,0} - \beta_{0,\pm1}}= 2\pi z_\mathrm{R}\approx 33.2~\milli\meter$, which agrees well with the oscillation period of the laser centroid shown in Fig. \ref{fig:mainresults}(b).

\begin{figure}
\centering
\includegraphics[width=1.0\linewidth]{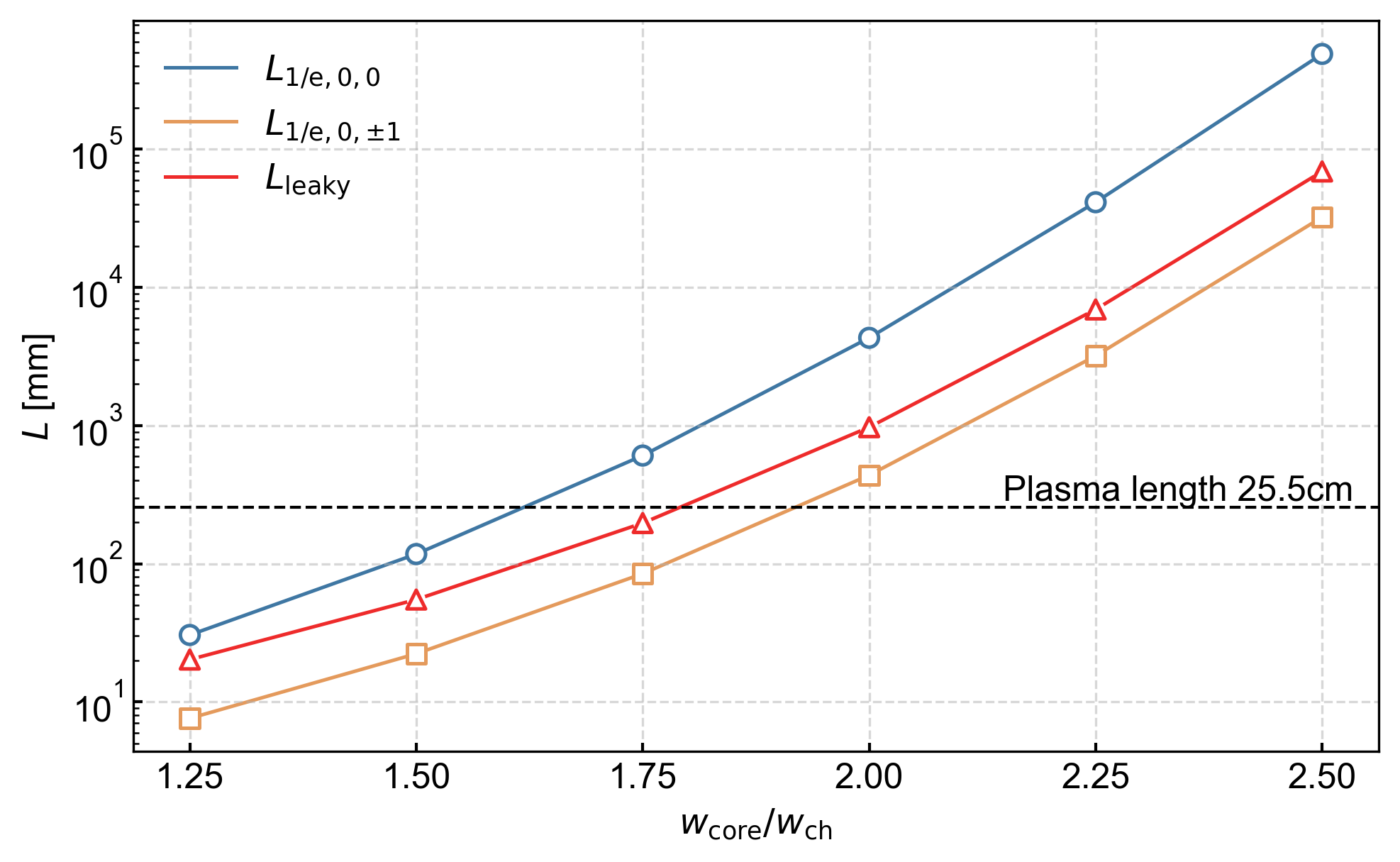}
\caption{\label{fig:L_leaky} Dependence of $L_{1/e,0,0}, L_{1/e,0,\pm1}$ and $L_\mathrm{leaky}$ on the channel parameter $w_\mathrm{core}$. The values of $L_{1/e,0,0}, L_{1/e,0,\pm1}$ are obtained by numerically solving the Helmholtz equation, while $L_\mathrm{leaky}$ is calculated as $L_\mathrm{leaky}=2\frac{L_{1/e,0,0}L_{1/e,0,\pm1}}{L_{1/e,0,0}-L_{1/e,0,\pm1}}$.}
\end{figure}

The corresponding eigenmodes of a finite channel can be known by numerically solving the Helmholtz equation \cite{Clark1998PhysRevLett.81.357, ClarkPhysRevE.61.1954}. We take the channel profile used in the simulations as an example and present its field profiles in the Appendix B. The field profiles of the fundamental and LG$_{0,\pm1}$ modes differ negligibly from those in an infinite channel. In contrast to the bounded modes of an infinite channel, the eigenmodes of a finite channel are inherently leaky, i.e., their intensities decrease as $I=I_0\exp\left(-\frac{z}{L_{1/e}}\right)$ \cite{Clark1998PhysRevLett.81.357, ClarkPhysRevE.61.1954}. Numerical calculations yield an $L_{1/e}$ of 4.32 m for the fundamental mode and 0.44 m for the LG$_{0,\pm1}$ mode. The difference in $L_{1/e}$ between these modes leads to an exponential decay of the centroid oscillation amplitude as $L_\mathrm{leaky}=2\frac{L_{1/e,0,0}L_{1/e,0,\pm1}}{L_{1/e,0,0}-L_{1/e,0,\pm1}}\approx 0.97~\meter$. This formula can be approximated as $L_\mathrm{leaky}\approx 2L_{1/e,0,\pm1}$ when $L_{1/e,0,0}\gg L_{1/e,0,\pm1}$ which is satisfied for most interested channels. 

We perform a scan over the channel parameter $w_\mathrm{core}$ while keeping all other parameters fixed, and present $L_{1/e,0,0}$ (blue), $L_{1/e,0,\pm1}$ (orange) and $L_\mathrm{leaky}$ (red) in Fig. \ref{fig:L_leaky}. They increase rapidly with increasing $w_\mathrm{core}$, consistent with the fact that a wider channel provides stronger confinement of the laser. For LWFA, a channel with $\frac{w_\mathrm{core}}{w_\mathrm{ch}} \gtrsim 2$ is essential to ensure that $\frac{L_{1/e,0,0}}{L_\mathrm{p}}\gg1$.

Since a laser pulse with a finite duration of $\tau_\mathrm{L}=42~\femto\second$ is considered here, the temporal walk-off of these two modes can also lead to a decrease of the centroid oscillation amplitude. It gives an characterized decay length of $L_\mathrm{w} \approx \frac{k_\mathrm{L}^2w_\mathrm{ch}^2c\tau_\mathrm{L}}{2} \approx 0.42~\meter$ \cite{Shrock2024PhysRevLett.133.045002}, derived from the group velocity of matched modes in an infinite channel, $v_{\mathrm{g},p,m}=\left( \frac{\partial \beta_{p,m}}{\partial \omega}\right)^{-1} = c\left[ 1-\frac{\omega_{\mathrm{p0}}^2}{2\omega_\mathrm{L}^2}-\frac{2(2p+|m|+1)}{k_\mathrm{L}^2 w_0^2}\right]$. $L_\mathrm{w}$ represents the propagation distance over which the slippage between the modes reaches one FWHM pulse duration $\tau_\mathrm{L}$. We also numerically calculate the propagation constant $\beta(\omega)$ near $\lambda_\mathrm{L}=1~\micro\meter$ based on the Helmholtz equation for a finite channel. The resulting group velocities of the fundamental and LG$_{0,\pm1}$ modes give a walk-off length of $0.41~\meter$, which is in excellent agreement with the formula. Since $L_\mathrm{w}$ scales linearly with the pulse duration, the decay of $\hat{x}_\mathrm{c}$ can be controlled by tuning the pulse duration.

\begin{figure}
\centering
\includegraphics[width=1.0\linewidth]{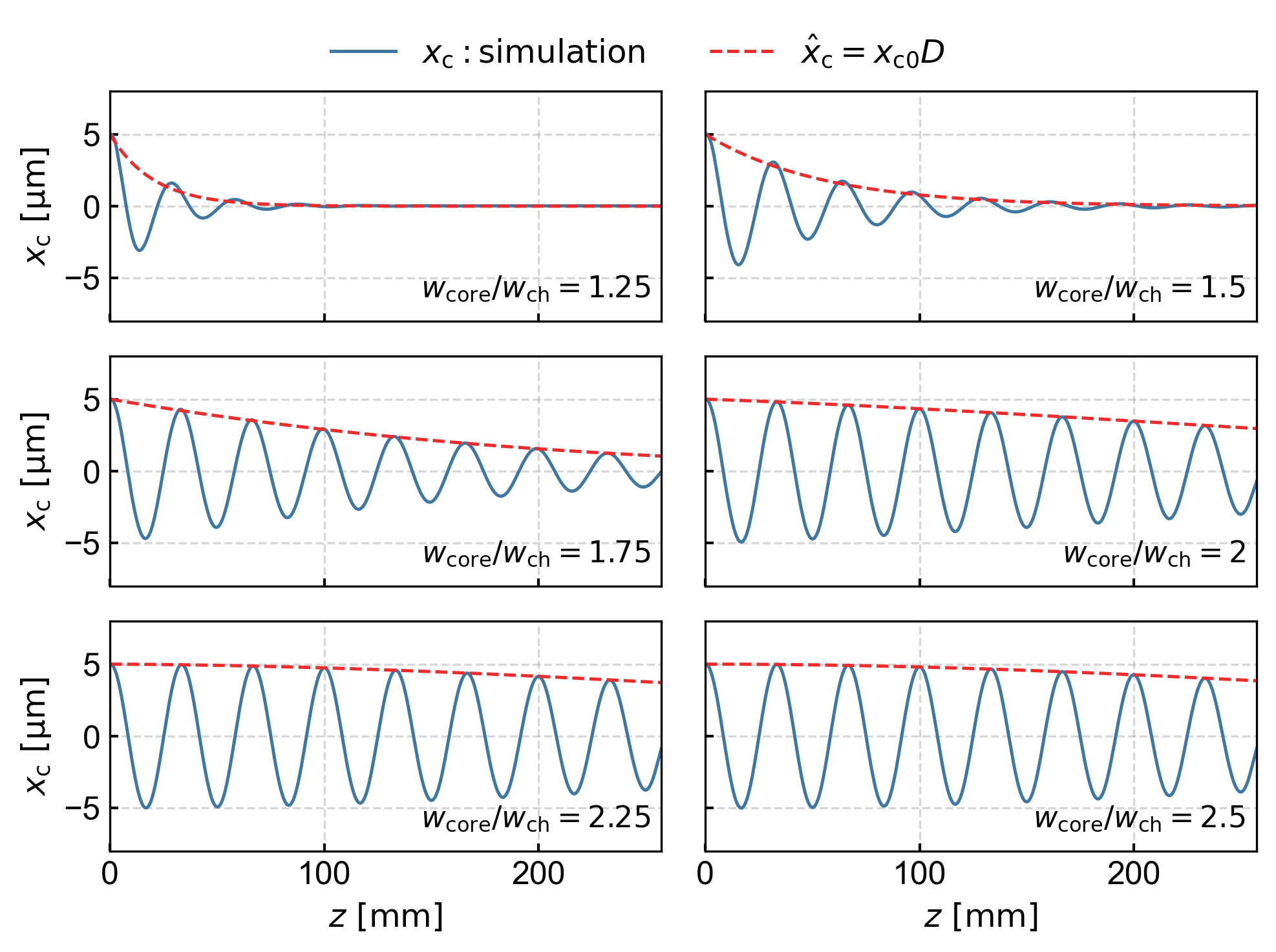}
\caption{\label{fig:weaklaser} The evolution of the non-relativistic laser centroid under different $w_\mathrm{core}$. The blue solid lines and red dashed lines represent the simulation results and analytical predication using Eq. \eqref{eq:D}, respectively. }
\end{figure}

We can combine the mode leakage and the walk-off to get a description of the decay of $\hat{x}_{c}$ for non-relativistic laser pulses. The offset laser electric field can be expressed as $E_\mathrm{L} \approx E_{0,0}f_{0,0}\mathrm{exp}(-\frac{z}{2L_{1/e,0,0}}) + \frac{x_\mathrm{c0}}{\sqrt{2}w_0}(E_{0,1} + E_{0,-1})f_{0,1}\mathrm{exp}(-\frac{z}{2L_{1/e,0,\pm1}})\mathrm{e}^{i\Delta\beta z}$, where $f(t-\frac{z}{v_{\mathrm{g},p,m}})$ is the temporal envelope, $\Delta\beta$ is the propagation constant difference between the fundamental and LG$_{0,\pm1}$ modes. Substituting the expression of the field into the definition of $x_\mathrm{c}$, we get \begin{align}
    x_\mathrm{c} \approx x_\mathrm{ci}\cos(\Delta \beta z)\frac{\int\mathrm{d}t f_{0,0} f_{0,1} }{\int\mathrm{d}t |f_{0,0}|^2}\exp\left( -\frac{z}{L_\mathrm{leaky}}\right),
\end{align} 
where the ratio term represents the contribution from the walk-off and the exponential term represents the contribution from the mode leakage. We use $D(z)= \frac{\int\mathrm{d}t f_{0,0} f_{0,1} }{\int\mathrm{d}t |f_{0,0}|^2}\exp\left( -\frac{z}{L_\mathrm{leaky}}\right)$ to represents the relative decay of the centroid oscillation amplitude. For a Gaussian temporal envelope $f(t)=\exp{\left(-2\ln{2}\frac{t^2}{\tau_\mathrm{L}^2}\right)}$, we can get
\begin{align}
    D(z) = \exp\left(-\ln2\frac{z^2}{L_{\mathrm w}^2} - \frac{z}{L_\mathrm{leaky}}\right). \label{eq:D}
\end{align}
The blue dashed line in Fig. \ref{fig:mainresults}(b) represents a fit to the simulation results using Eq. \eqref{eq:D}. It gives $L_\mathrm{leaky}\approx 0.94~\meter$ and $L_\mathrm{w}\approx 0.42~\meter$, which aggress well with the theoretical analysis.

We perform a series of 3D simulations to scan $w_\mathrm{core}$ and show the evolution of $x_\mathrm{c}$ for non-relativistic laser pulses in Fig. \ref{fig:weaklaser}. Their envelopes all agree well with Eq. \eqref{eq:D} (red dashed lines). When $\frac{w_\mathrm{core}}{w_\mathrm{ch}}$ is small (e.g., $\frac{w_\mathrm{core}}{w_\mathrm{ch}}=1.25$ and 1.5), the decay of $\hat{x}_\mathrm{c}$ is dominated by mode leakage and exhibits an exponential behavior. After a few oscillation cycles, most of the energy in the higher-order modes is radiated away through the channel boundary. This may be utilized to stabilize the laser propagation. In contrast, when $\frac{w_\mathrm{core}}{w_\mathrm{ch}}$ is large (e.g., $\frac{w_\mathrm{core}}{w_\mathrm{ch}}=2.25$ and 2.5), the decay is dominated by walk-off and shows a Gaussian behavior. The higher-order modes gradually detach from the fundamental mode with negligible energy loss. For intermediate values, the decay is governed by a combination of both effects. 

\begin{figure*}
\centering
\includegraphics[width=0.75\linewidth]{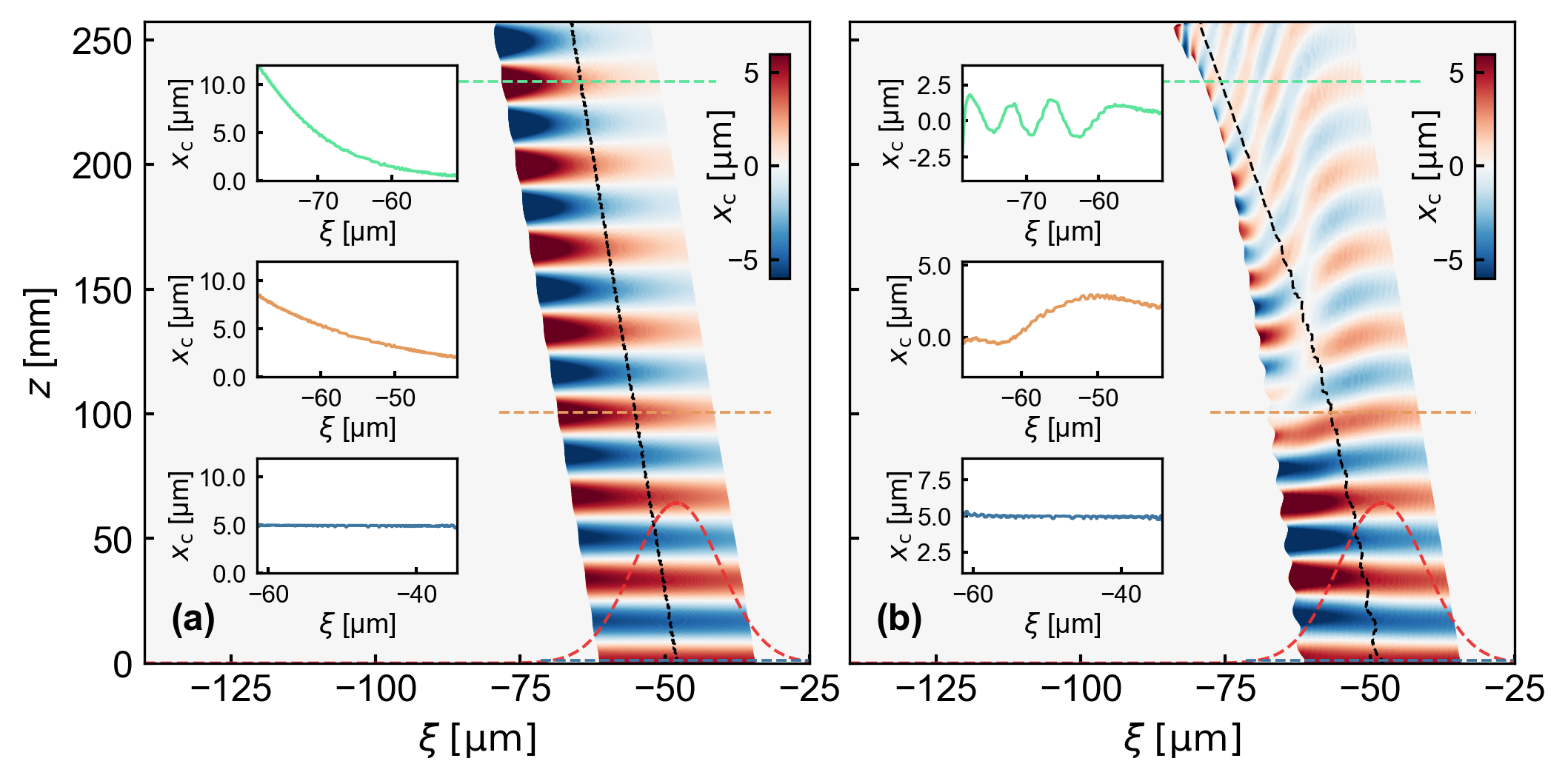}
\caption{\label{fig:laser-centroid-map} Laser slice centroid evolution for $a_0=0.01$ (a) and $a_0=2.5$ (b). The black dashed lines represent the location of the peak $a_0$. The red dashed lines represent the initial laser intensity profile. The insets in each panel show the slice centroid at three representative distances as indicated by the colored dashed lines ($z$=1.0, 100.6 and 233.3 mm).}
\end{figure*}

\section{IV. Laser centroid oscillation in the relativistic case}
When a relativistic laser pulse propagates in a preformed plasma channel, it modifies the channel profile through its ponderomotive force and the plasma wake driven by the front part can act on the trailing part \cite{Benedetti2012PoP}. Meanwhile, the self-phase modulation and the pump depletion can significantly alter the spatiotemporal and spectral profiles of the laser \cite{mori1997physics}. These combined effects lead to a centroid oscillation behavior that differs from that in the non-relativistic case, as shown in Fig. \ref{fig:mainresults}(b). We present an analysis of the slice-dependent laser centroid for both cases in Fig. \ref{fig:laser-centroid-map}, where $\xi\equiv z-ct$ is the co-moving coordinate. For the non-relativistic case [(a)], different axial slices of the laser oscillate at the same frequency. Due to the walk-off, the LG$_{0,\pm1}$ mode slips backward relative to the fundamental mode. Consequently, the field ratio between the LG$_{0,\pm1}$ and fundamental modes decreases in the leading slices while increasing in the trailing slices. As shown in Fig. \ref{fig:laser-centroid-map}(a) and its three insets, this manifests as a decrease in $\hat{x}_\mathrm{c}$ at the leading slices and an increase at the trailing slices. For the relativistic laser case [Fig. \ref{fig:laser-centroid-map}(b)], the slice centroid behavior is similar to that of the non-relativistic case before $z\approx 70~\milli\meter$. Subsequently, the centroid oscillation period at the trailing slices gradually decreases, while that at the leading slices changes little. This axially varying period builds an oscillation phase difference along the laser pulse. As shown by the middle inset, the slice centroid oscillates along the axial direction and the number of the oscillation period increases with propagation distance (see the top inset). The mixing of the centroid oscillation phase \cite{PhysRevLett.112.035003} causes the fast decay of the overall laser centroid as shown in Fig. \ref{fig:mainresults}(b).

\begin{figure}
\centering
\includegraphics[width=1.0\linewidth]{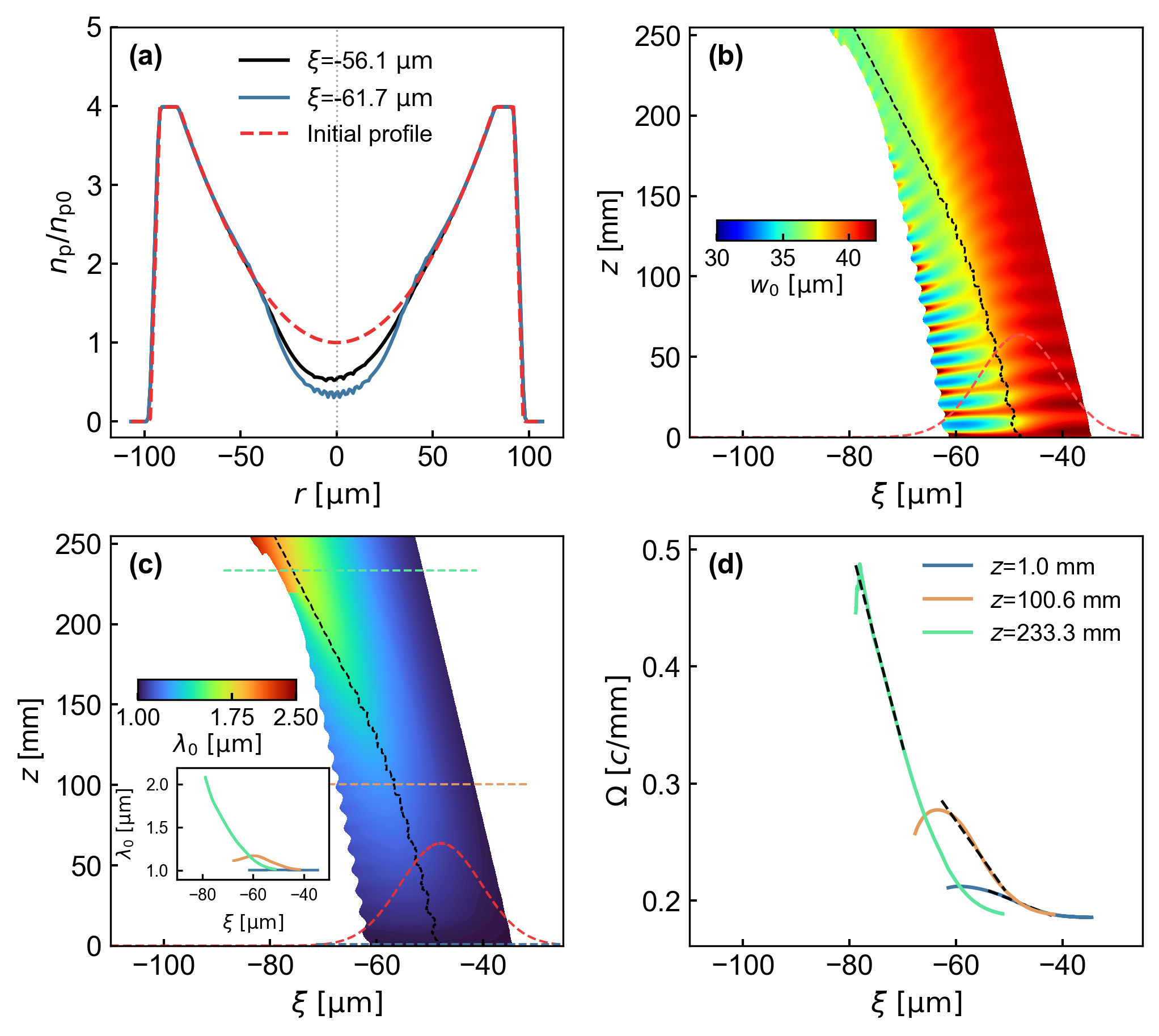}
\caption{\label{fig:frequency-map} The centroid dynamics in the relativistic case ($a_0=2.5$). (a) The plasma electron density distribution at two slices at a propagation distance of 25.5 mm (black and blue solid lines) and the initial profile is shown by the red dashed line. (b) The evolution of the laser spot size at each slice. (c) The evolution of the central wavelength at each slice. (d) The slice oscillation frequency at three representative distances. The black and red dashed lines in (b) and (c) represent the location of the peak $a_0$ and the initial laser profile, respectively. }
\end{figure}

A relativistic laser pulse can push the electrons outward from the channel center through its ponderomotive force and modify the electron density profile of the channel. Generally, the laser deepens the channel by reducing the on-axis plasma electron density. The black and blue lines in Fig. \ref{fig:frequency-map}(a) show the transverse profiles of the plasma electron density at two trailing slices ($\xi=-56.1$ and -61.7 $\micro\meter$) at a laser propagation distance of 25.5 mm. The on-axis density is reduced relative to the initial value (indicated by the red dashed line), with the further trailing slice exhibiting an even lower density. The relativistic mass effect further increases the refractive index difference between the axis and the channel boundaries \cite{mori1997physics}. The slice-matched spot size of the modified channel [$w_\mathrm{m}(\xi)$] decreases from the initial $w_\mathrm{ch}$ at the very front of the laser pulse \cite{Benedetti2012PoP}. Figure \ref{fig:frequency-map}(b) shows the evolution of the laser spot size at each slice. The leading slices are approximately matched with the channel and the spot size remain close to the initial value (41 $\micro\meter$). Meanwhile, the spot sizes of the trailing slices undergo substantial oscillations with an amplitude of $\approx 6~\micro\meter$.

In the nonlinear LWFA, a co-moving refractive index gradient along the propagation direction is created by the laser ponderomotive force, which can decelerate the laser photons \cite{mori1997physics, LuPoP2025-1, LuPoP2025-2}. This can be reflected by calculating the central wavelength along the laser. Figure \ref{fig:frequency-map}(c) illustrates the evolution of the central wavelength. A significant photon deceleration occurs after $z\approx 50~\milli\meter$. As shown in the inset, a wavelength chirp is developed along the laser, i.e., the photons in the trailing slices are decelerated more, and the chirp grows as the laser propagates. 

Since the centroid oscillation frequency $\Omega\equiv \frac{c}{z_\mathrm{R}}$ is determined by the matched spot size and the laser wavelength, the axially varying $w_\mathrm{m}(\xi)$ and $\lambda_\mathrm{L}(\xi)$ result in a slice-dependent $\Omega(\xi)$. We show the axial profile of $\Omega$ at three representative distances in Fig. \ref{fig:frequency-map}(d). At each $z$, we find the nearby maximum and minimum of the spot size at each laser slice and use their average value as the matched spot size. At the beginning ($z=1.0~\milli\meter$), the laser wavelength remains at the initial value [blue line in the inset of Fig. \ref{fig:frequency-map}(c)] and the axial $\Omega$ variation is dominated by $w_\mathrm{m}(\xi)$, with $\Omega$ increasing towards the laser tail as shown by the blue line in Fig. \ref{fig:frequency-map}(d). As the photon deceleration occurs, the laser wavelength gradually decreases, and the resulting wavelength chirp [brown line in the inset of Fig. \ref{fig:frequency-map}(c)] increases the axial variation of $\Omega$ [brown line in the inset of Fig. \ref{fig:frequency-map}(d)]. At $z=100.6~\milli\meter$, the phase difference between the laser head and tail accumulates to $\sim 1.5\pi$ [brown line in Fig. \ref{fig:laser-centroid-map}(b)] and causes the decrease of the overall centroid oscillation. The photon deceleration continues and results in a rapid decay of the overall centroid oscillation. As shown by the black dashed lines in \ref{fig:frequency-map}(d), the chirp of $\Omega$ around the laser peak intensity grows from $\frac{\mathrm{d}\Omega}{\mathrm{d}\xi}\approx -0.0018~c/(\milli\meter\cdot\micro\meter)$ at $z=1.0~\milli\meter$ to $\approx -0.0064~c/(\milli\meter\cdot\micro\meter)$ at $z=100.6~\milli\meter$ and $\approx -0.0175~c/(\milli\meter\cdot\micro\meter)$ at $z=233.3~\milli\meter$.

To quantify the effect of the phase mixing on the centroid decay, we provide a toy model by assuming a non-evolving Gaussian temporal laser envelop and a linear and non-evolving chirp of the oscillation frequency as $\Omega(\xi^\prime)=\Omega_\mathrm{mid} + \kappa \xi^\prime$, where $\xi^\prime=z- v_g t$ is the coordinate moving with the laser pulse and $\Omega_\mathrm{mid}$ is the value at the middle of the pulse. For a Gaussian intensity envelope, the overall centroid motion can be integrated as $x_{\mathrm{c}}=x_\mathrm{ci} \frac{\int \mathrm{d}\xi\cos[\Omega(\xi)z/c]I(\xi)}{\int\mathrm{d}\xi I(\xi)} = x_\mathrm{ci}\exp\left(-\frac{z^2}{2 L_\mathrm{PM}^2}\right) \cos\left(\frac{\Omega_\mathrm{mid}}{c} z \right)$, where $L_\mathrm{PM}=\frac{2\sqrt{2\ln{2}}}{\lvert\kappa\rvert \tau_\mathrm{L}}$ is the characteristic decay length caused by phase mixing. As governed by the Gaussian profile, a rapid decay in $\hat{x}_\mathrm{c}$ occurs once $z \gtrsim L_\mathrm{PM}$. This agrees with the behavior of the laser centroid for $a_0=2.5$ as shown by the orange line in Fig. \ref{fig:mainresults}(b). A Gaussian fitting of $\hat{x}_\mathrm{c}$ gives $L_\mathrm{PM}\approx 86~\milli\meter$, which corresponds to $\lvert\kappa\rvert\approx 0.0022~c/(\milli\meter\cdot\micro\meter)$ for $\tau_\mathrm{L}=42~\femto\second$. It aggress reasonable with the chirp observed in the simulations. An accurate description of the $\hat{x}_\mathrm{c}$ requires a self-consistent treatment of the spatiotemporal and spectral distributions of the laser pulse.

%\textcolor{blue}{An interesting point: The macroscopic centroid amplitude decays most rapidly at the inflection point of the Gaussian envelope, $z_{\mathrm{inflect}} = 2/(|\kappa|\sigma)$. At this critical propagation distance, the accumulated phase difference across the effective pulse length ($2\sigma$) reaches $\Delta \Phi = 4$ rad ($\approx 1.27\pi$), triggering a regime of maximal destructive interference where the out-of-phase longitudinal slices violently cancel each other. As indicated by the orange line in Fig. 4(b), this point corresponds to a phase difference of approximately $1.2\pi$ and to the location of the most rapid decay in Fig. 6 across all five cases.}

\begin{figure}
\centering
\includegraphics[width=1.0\linewidth]{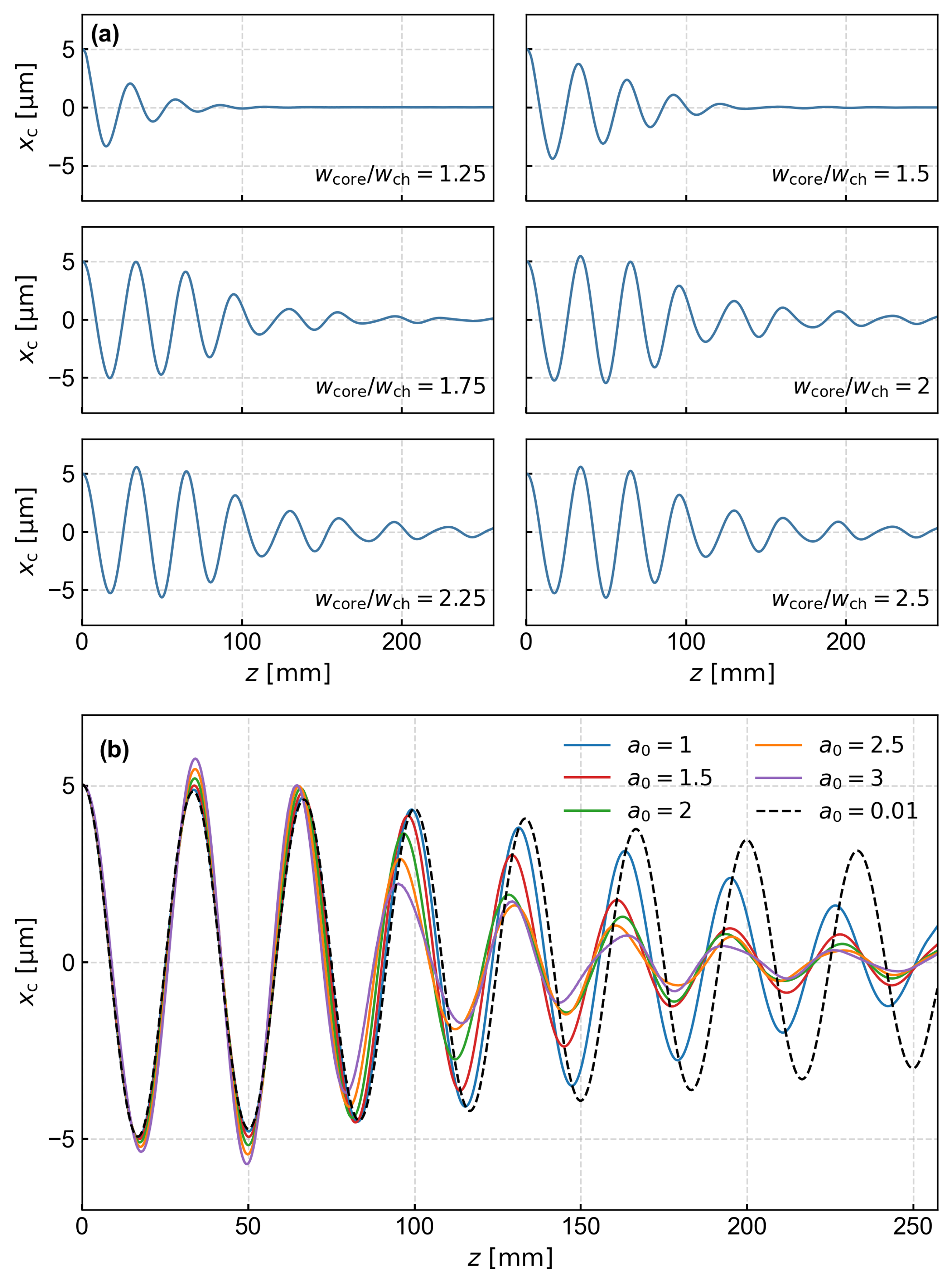}
\caption{\label{fig:scan} The evolution of the relativistic laser centroid under different $w_\mathrm{core}$ (a) and $a_0$ (b). $a_0=2.5$ in all panels of (a) and $\frac{w_\mathrm{core}}{w_\mathrm{ch}}=2$ in (b). The non-relativistic laser case is also presented in (b) for comparison.}
\end{figure}

We perform a parameter scan over the channel parameter $w_\mathrm{core}$ for $a_0=2.5$ and present the centroid oscillation in Fig. \ref{fig:scan}(a). When $\frac{w_\mathrm{core}}{w_\mathrm{ch}}\gtrsim2$, the mode leakage and the walk-off induce only a minor decay of $\hat{x}_\mathrm{c}$ as shown in Fig. \ref{fig:weaklaser}. Thus, the phase mixing dominates the decay for relativistic lasers, which is evidenced by the similar $x_\mathrm{c}$ evolution for cases with $\frac{w_\mathrm{core}}{w_\mathrm{ch}}=2$, 2.25 and 2.5. For narrower channels ( $\frac{w_\mathrm{core}}{w_\mathrm{ch}}=1.25$ and 1.5), the mode leakage reduces the $\hat{x}_\mathrm{c}$ rapidly, which dominates its overall behavior. For intermediate values of $\frac{w_\mathrm{core}}{w_\mathrm{ch}}$ (e.g., 1.75), these effects mutually determine the $\hat{x}_\mathrm{c}$ evolution. 

Since the rapid decay in the relativistic laser case is driven by the phase mixing, which is in turn dependent on the laser intensity, we scan $a_0$ from 1 to 3 and show the centroid behavior in Fig. \ref{fig:scan}(b). As $a_0$ increases, the plasma channel is modified more significantly and photon deceleration develops more rapidly, leading to a faster decay of $\hat{x}_\mathrm{c}$. Furthermore, as $a_0$ grows, the oscillation period exhibits a more pronounced decrease compared to the nonrelativistic case (black dashed line). The initial slight increase of $\mathrm{x}_c$ (around $z=40~\milli\meter$) may be caused by the hosing effect \cite{Sprangle1994PhysRevLett.73.3544}, which is stronger for a larger $a_0$.

\section{V. Conclusions}
The propagation of short and intense laser pulses in preformed plasma channels is essential for various applications, particularly for driving high-energy and high-quality LWFAs. In contemporary high-power laser facilities, inherent system jitters inevitably cause the laser pulses to enter the channel with transverse position and/or angular offsets. In this work, we have investigated the centroid oscillation of such offset lasers propagating in plasma channels. For non-relativistic lasers, we find that the mode leakage in a finite-width channel and the temporal walk-off of a finite-duration pulse result in the decay of the centroid oscillation amplitude. For relativistic laser pulses, photon deceleration coupled with channel modification induces slice-dependent centroid oscillation behaviors. The phase mixing of the centroids among different axial slices leads to a rapid decay of the overall laser centroid. The phase mixing should also contribute to the experimentally observed decay of the laser spot size oscillation when a relativistic laser is injected into the channel with a mismatched size \cite{Miao2022PhysRevX.12.031038, Picksley2023PhysRevLett.131.245001, Shrock2024PhysRevLett.133.045002}. A comprehensive understanding of laser evolution under these realistic conditions is of great significance for optimization of LWFAs and plasma undulators.

\section*{Acknowledgments}
\begin{acknowledgments}
This work was supported by the Fundamental and Interdisciplinary Disciplines Breakthrough Plan of the Ministry of Education of China-JYB2025XDXM204, the National Grand Instrument Project No. SQ2019YFF01014400, National Natural Science Foundation of China No. 12375147, Beijing outstanding young scientist project, and the Fundamental Research Funds for the Central Universities, Peking University. The simulations were supported by the High-performance Computing Platform of Peking University.
\end{acknowledgments}

\section*{Data Availability}

The data that support the findings of this study are available from the corresponding author upon reasonable request.

\section*{Appendix A: PIC simulation setup}
The simulations are carried out using the fully relativistic, 3D electromagnetic PIC code OSIRIS \cite{fonseca2002high}. To mitigate computational costs, the simulations are performed in a Lorentz-boosted frame with a relativistic factor of $\gamma_\mathrm{b} = 10$ \cite{yu2014modeling, yu2016enabling}. Mapped to the laboratory frame, a simulation window moving at the speed of light has dimensions of $200~\micro\meter \times 214~\micro\meter \times 214~\micro\meter$ with $6400 \times 214 \times 214$ cells in the $z$, $x$, and $y$ directions, respectively. This corresponds to a spatial resolution of $\Delta z = 0.03125~\micro\meter$ (i.e., $\lambda_\mathrm{L}/32$) and $\Delta x = \Delta y = 1~\micro\meter$, which provides sufficient resolution to fully resolve the laser wavelength and accurately capture the transverse centroid offsets. To effectively suppress numerical noise related with relativistically drifting plasma \cite{xu2013numerical, xu2020numerical}, a customized Maxwell solver is employed \cite{xu2020numerical, li2021new}. There is 1 macroparticle per cell to represent the plasma electrons and ions. The background plasma starts at $z = 0~\milli\meter$ with a $100~\micro\meter$ longitudinal density upramp, followed by a $25.5~\centi\meter$ long uniform plateau, and a $100~\micro\meter$ density downramp.

\section*{Appendix B: mode analysis for a finite channel}
For a given channel profile, the eigenmodes are determined by the Helmholtz equation \cite{ClarkPhysRevE.61.1954},
\begin{equation}
\frac{d^{2}\mathcal{E}}{ds^{2}}+\frac{1}{s}\frac{d\mathcal{E}}{ds}
+\left[n^{2}(s)-\frac{\beta^{2}}{k_\mathrm{L}^{2}}-\frac{m^{2}}{s^{2}}\right]\mathcal{E}=0, \label{Helmholtz}
\end{equation}
where $E(r,z)=\mathcal{E}(r)e^{i\beta z}$ is the electric field, $s=k_\mathrm{L} r$ is the dimensionless radial coordinate, $n(s)$ is the radial refractive-index profile, $\beta$ is the propagation constant, and $m$ is the azimuthal index. All supported modes of a finite channel are intrinsically leaky and tunnel through the index barrier into the outer region. The refractive-index profile $n(r)$ of a finite channel is usually generally too complicated to allow an analytical solution of the Helmholtz equation, numerical solutions are routinely employed to determine the eigenmodes \cite{MiaoPhysRevLett.125.074801, Miao2022PhysRevX.12.031038, Shrock2024PhysRevLett.133.045002}. 

\begin{figure}
\centering
\includegraphics[width=1.0\linewidth]{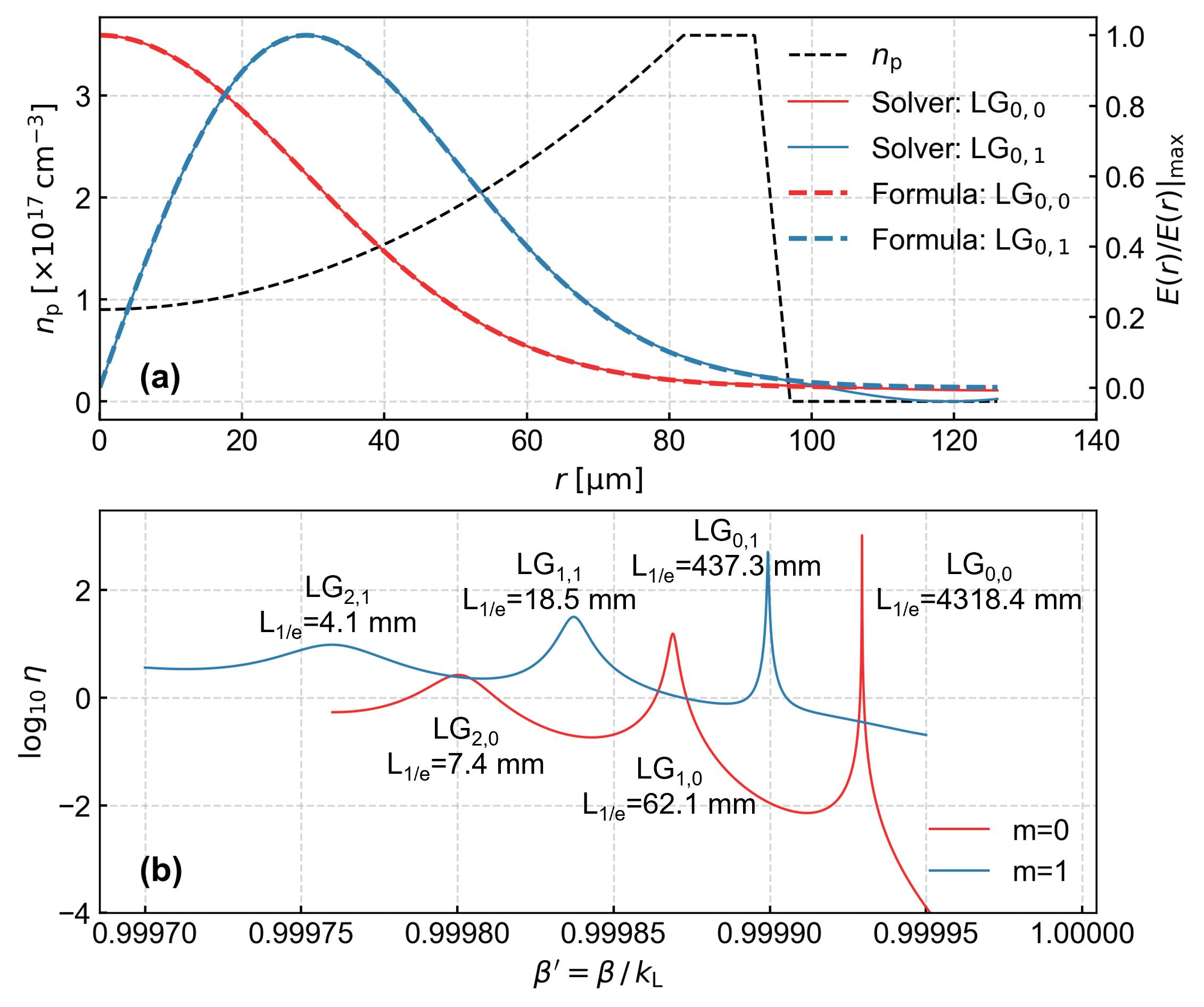}
\caption{\label{fig:mode} Numerical solutions of the Helmholtz equation. (a) The normalized field profile of the mode LG$_{0,0}$(red solid line) and mode LG$_{0,1}$ (blue solid line) for a finite channel whose density is shown by the black dashed line. The field profiles for a ideal channel are presented by the colored dashed lines as a comparison. Note that this channel is the same as that used in Fig. \ref{fig:mainresults}. (b) The spectrum $\eta(\beta')$ for $m=0$ (red) and $m=1$ (blue). The $1/e$ decay length is labeled for different modes.}
\end{figure}

To quantify the transverse confinement of these quasi-bound modes, we define the energy confinement ratio $\eta(\beta')=\frac{\int_A |\mathcal{E}(s)|^2\,dA}{A|\mathcal{E}_{\mathrm{vacuum}}|^2}$, where $\beta'=\beta/k_\mathrm{L}$ is the normalized propagation constant. By scanning $\beta'$ and integrating the radial equation outward from $s=0$, we obtain the spectrum $\eta(\beta')$ shown in Fig.~\ref{fig:mode}(b) for a channel profile as shown by the dashed line in Fig. \ref{fig:mode}(a). 

The discrete modes, labeled by $(p,m)$, correspond to the resonance peaks of $\eta(\beta')$. The normalized field profile for the fundamental mode (red solid line) and the LG$_{01}$ mode (blue solid line) are shown in Fig. \ref{fig:mode}(a). They change little compared with the modes of an ideal channel (colored dashed lines). The leakage rates of the modes are characterized by the width of the $\eta(\beta')$ peaks. Fitting each peak with a Lorentzian profile yields the full width at half maximum $\Delta\beta'$, from which the corresponding $1/e$ attenuation length is estimated as $L_{1/e}=\frac{1}{k_\mathrm{L}\Delta\beta'}$. Broad peaks indicate strong radiative leakage, while narrow peaks correspond to well-confined quasi-bound modes with longer propagation lengths.

\bibliographystyle{apsrev4-1}
%reverse

%\raggedright
\bibliography{refs_xinlu}
\end{document}